\documentclass[aps,prd,twocolumn,nofootinbib]{revtex4-1}
\usepackage{epsfig}
\usepackage[colorlinks,linkcolor=blue,anchorcolor=blue,citecolor=blue,urlcolor=blue,breaklinks=true]{hyperref}
\usepackage{graphics}
\usepackage{color}

\begin{document}
\author{Pei-Lin Yin$^{1,4}$}
\author{Hai-Xiao Xiao$^{2}$}
\author{Hong-Tao Feng$^{1,4}$}\email{Email:fenght@seu.edu.cn}
\author{Hong-Shi Zong$^{2,3,4}$}\email{Email:zonghs@nju.edu.cn}

\address{$^{1}$Department of Physics, Southeast University, Nanjing 211189, China}
\address{$^{2}$Department of Physics, Nanjing University, Nanjing 210093, China}
\address{$^{3}$Joint Center for Particle, Nuclear Physics and Cosmology, Nanjing 210093, China}
\address{$^{4}$State Key Laboratory of Theoretical Physics, Institute of Theoretical Physics, CAS, Beijing 100190, China}

\title{Effects of gauge boson mass on chiral and deconfinement phase transitions in QED$_{3}$}
\begin{abstract}
Based on the experimental observation that there is a coexisting region between the antiferromagnetic (AF) and $\textit{d}$-wave superconducting ($\textit{d}$SC) phases, the influences of gauge boson mass $m_{a}$ on chiral symmetry restoration and deconfinement phase transitions in QED$_{3}$ are investigated simultaneously within a unified framework, i.e., Dyson-Schwinger equations. The results show that the chiral symmetry restoration phase transition in the presence of the gauge boson mass $m_{a}$ is a typical second-order phase transition; the chiral symmetry restoration and deconfinement phase transitions are coincident; the critical number of fermion flavors $N^{c}_{f}$ decreases as the gauge boson mass $m_{a}$ increases and there exists a boundary that separates the $N^{c}_{f}$-$m_{a}$ plane into chiral symmetry breaking/confinement region for ($N_{f}^{c}$, $m_{a}$) below the boundary and chiral symmetry restoration/deconfinement region for ($N_{f}^{c}$, $m_{a}$) above it.
\bigskip

\noindent Keywords: QED$_3$, Chiral and deconfinement phase transitions, Dyson-Schwinger equations
\bigskip

\noindent PACS Number(s):  11.10.Kk, 11.15.Tk, 11.30.Qc
\end{abstract}
\maketitle

\section{Introduction}
Dynamical chiral symmetry breaking (DCSB) and quark color confinement are two fundamental features of Quantum chromodynamics (QCD) that describes the interactions between quarks and gluons. Research on these two nonperturbative phenomena is conducive to deepening our understanding of the nature as well of the early Universe and, thus, becomes one of the central themes in today's theoretical calculations and experimental measurements. In the past forty years, a great many efforts have been devoted in this field~\cite{Phys.Rev.D.58.096007,Phys.Rev.Lett.86.592,Nucl.Phys.B.673.170,Phys.Rep.407.205,Nucl.Phys.A.772.167,
Nature.443.675,Rev.Mod.Phys.80.1455,Rep.Prog.Phys.74.014001,Phys.Rev.Lett.106.172301,Phys.Rev.D.85.054503,Prog.Theor.Phys.Suppl.193.1,J.High.Energy.Phys.07.014,
Ann.Phys.358.172}. However, because of the complicated non-Abelian feature of QCD itself, it is difficult to have a thorough understanding of the mechanisms of DCSB and confinement. In this case, to gain a valuable insight into them, it is very suggestive to study some models which are structurally much simpler than QCD while sharing the same basic nonperturbative phenomena.

Quantum electrodynamics in (2+1)-dimensions (QED$_{3}$) is just such a model. It has several nonperturbative features similar to QCD, such as asymptotic freedom~\cite{Phys.Rev.D.33.3704,Prog.Theor.Phys.87.193,Phys.Rev.D.55.7826,Prog.Theor.Phys.105.809,arXiv:1410.0118}, DCSB~\cite{Phys.Lett.B.253.246,
Phys.Lett.B.266.163,Phys.Lett.B.295.313,Phys.Rev.D.50.1068,Phys.Rev.D.58.105012,J.High.Energy.Phys.03.020,Phys.Lett.B.491.280,Phys.Rev.D.68.025017,
Mod.Phys.Lett.A.22.449,Phys.Rev.D.81.045006,Phys.Rev.D.84.036013,Phys.Rev.D.86.065002,Ann.Phys.348.306}, and confinement~\cite{Phys.Rev.D.46.2695,
Phys.Rev.D.52.6087,J.High.Energy.Phys.09.048,Phys.Rev.B.79.014507,Phys.Rev.D.82.067701}. In addition, due to the coupling constant being dimensionful (its dimension is $\sqrt{mass}$), QED$_{3}$ is superrenormalizable and does not suffer from the ultraviolet divergences which are present in the corresponding four-dimensional theories. These properties of QED$_{3}$ enable it to serve as a toy model of QCD. In parallel with its relevance as a tool through which to develop insight into aspects of QCD, QED$_{3}$ is also of interest in condensed matter physics as a low-energy effective field theory for strongly correlated electronic systems, such as high-$T_{c}$ cuprate superconductors~\cite{Phys.Rev.Lett.79.2109,Ann.Phys.272.130,Phys.Rev.Lett.86.3871,
Phys.Rev.Lett.87.257003,Phys.Rev.B.65.180511,Phys.Rev.B.66.054535,Phys.Rev.B.66.094504,Phys.Rev.Lett.88.047006,Rev.Mod.Phys.78.17} and graphene ~\cite{Phys.Rev.Lett.95.146801,PhysRevLett.96.256802,Phys.Rev.B.79.205429,Phys.Rev.Lett.102.026802}.

It is well known that the Lagrangian of the system holds on chiral symmetry in the chiral limit and the chiral symmetry of the vacuum will be broken dynamically when the massless fermion acquires a nonzero mass through nonperturbative effects. The breakthrough in research into DCSB in QED$_{3}$ with $N_{f}$ flavors of massless fermions was finished by T.W. Appelquist $\textit{et al.}$~\cite{Phys.Rev.Lett.60.2575}. They first solved numerically the Dyson-Schwinger equation (DSE) for the fermion self-energy function up to the leading-order in 1/$N_{f}$ expansion and concluded that DCSB takes place only when the number of fermion flavors $N_{f}$ is less than a critical number of fermion flavors $N_{f}^{c}$=32/$\pi^{2}$. After taking into account the next-to-leading-order corrections to the fermion wave-function renormalization, D. Nash~\cite{Phys.Rev.Lett.62.3024} then obtained $N_{f}^{c}$=128/3$\pi^{2}$. Later, P. Maris~\cite{Phys.Rev.D.54.4049} solved a set of coupled integral equations for the fermion wave-function renormalization, the fermion self-energy function, and the boson vacuum polarization with a range of simplified fermion-boson vertices and arrived at $N_{f}^{c}$$\approx$3.3. C. S. Fisher $\textit{et al.}$~\cite{Phys.Rev.D.70.073007} employed power laws for the fermion wave-function renormalization and boson vacuum polarization to investigate the infrared behavior of the coupled system of fermion and boson equations and found $N_{f}^{c}$$\approx$4. A. Bashir $\textit{et al.}$~\cite{Phys.Rev.C.78.055201} employed an efficacious models for the boson vacuum polarisation and fermion-boson vertex and yielded a value $N_{f}^{c}$$\approx$3.24. Recently, J. Braun $\textit{et al.}$~\cite{Phys.Rev.D.90.036002} studied the many-flavor phase diagram of QED$_{3}$ by analyzing the RG fixed-point structure of the theory and found that the phase transition towards a chirally broken phase occurring at small flavor numbers could be separated from the quasiconformal phase at larger flavor numbers by an intermediate phase.

The above result holds when the gauge boson is massless, but it is expected to change as the gauge boson acquires a finite mass $m_{a}$. DCSB is a low-energy nonperturbative phenomenon and realized by forming fermion-antifermion condensation mediated by a strong long-range gauge interaction. However, when the gauge boson has a finite mass $m_{a}$, the gauge interaction between fermions is significantly weakened. Intuitively, a finite gauge boson mass $m_{a}$ is repulsive to DCSB. On the other hand, in some high-$T_{c}$ superconducting experiments, such as neutron scattering~\cite{Nature.415.299,Science.291.1759}, muon-spin resonance ($\mu$SR)~\cite{Phys.Rev.Lett.88.137002}, and scanning tunneling microscopy~\cite{Science.295.466} experiments, it has been found that there is a region of the coexistence of antiferromagnetic (AF) phase (in which the fermion acquires a nonzero mass by DSCB while the gauge boson remains massless~\cite{Phys.Rev.Lett.87.257003,Phys.Rev.B.65.180511,Phys.Rev.Lett.88.047006}) and $\textit{d}$-wave superconducting ($\textit{d}$SC) phase (where the gauge boson has a finite mass $m_{a}$ via the Anderson-Higgs mechanism, but the fermion becomes massless~\cite{Phys.Rev.Lett.87.257003,Phys.Rev.B.65.180511,Phys.Rev.Lett.88.047006}). This phenomenon implies a certain interplay between the AF phase and the $\textit{d}$SC phase, which is one of the fundamental issues in modern condensed matter physics. Thus it is very interesting to study the effect of gauge boson mass $m_{a}$ on DCSB.

It is commonly believed that DCSB and confinement are nonperturbative phenomena that have to be studied in a nonperturbative way. The Dyson-Schwinger equations (DSEs) provide a natural framework within which to explore these and related phenomena. In this paper, we will investigate the influences of gauge boson mass $m_{a}$ on chiral symmetry restoration and deconfinement phase transitions simultaneously within a unified framework, i.e., DSEs. The paper is organized as follows: In Sec. II, we introduce two criteria (the chiral condensate and the Schwinger function) that are used to characterize the nature of the chiral symmetry restoration and the deconfinement phase transitions, respectively. In Sec. III, the DSEs satisfied by three scalar functions of the fermion and boson propagators are set up. In Sec. IV, we solve numerically the DSEs for the fermion and boson propagators and then calculate the chiral condensate and the function $\kappa(N_{f})$ within a range of the numbers of fermion flavors $N_{f}$ and the gauge boson mass $m_{a}$. A brief summary and discussion are given in Sec. IV.

\section{Criteria For The Chiral And Deconfinement Phase Transitions}
The chiral condensate is the vacuum expectation value of scalar operator $\bar\psi\psi$. The nonzero value of which indicates that the chiral symmetry reflected on the Lagrangian level is spontaneously broken on the vacuum level and the chiral symmetry gets restored when it vanishes for the chiral limit, which makes it possible to define the chiral condensate as the order parameter for the chiral symmetry restoration phase transition. The chiral condensate can be obtained by differentiating the generating functional with respect to the current mass of the fermion and further expressed in terms of the dressed fermion propagator by means of functional analysis method
\begin{eqnarray}
\langle\bar\psi\psi\rangle(N_{f})=-\int\frac{\textrm{d}^3p}{(2\pi)^3}\textrm{Tr}[S(p,N_{f})]=-\frac{\partial \ln Z}{\partial m},\label{eq1}
\end{eqnarray}
where the notation Tr denotes the trace operation over Dirac indices of the dressed fermion propagator $S(p)$.

Dynamical properties of a many-particle system can also be investigated by measuring the response of the system to an external perturbation that disturbs the system only slightly in its equilibrium state. In previous studies related to the chiral symmetry restoration phase transition, a widely used response function is chiral susceptibility~\cite{Phys.Rev.D.50.6954,Phys.Lett.B.591.277,Phys.Rev.C.72.035202,Phys.Lett.B.643.46,Phys.Rev.D.77.076008,Phys.Rev.C.79.035209} that is defined as the first-order derivative of the order parameter (i.e., the chiral condensate) with respect to the current mass of the fermion. Because the chiral condensate behaves differently in chiral symmetry breaking and restoration phases, the chiral susceptibility often exhibits some singular behaviors, such as discontinuity or divergence, which are usually regarded as essential characteristics of the chiral phase transition. By definition, the chiral susceptibility is written as
\begin{eqnarray}
\chi^{c}(N_{f})=-\frac{\partial\langle\bar\psi\psi\rangle(N_{f})}{\partial m}\bigg|_{m\rightarrow0}=\frac{\partial^{2} \ln Z}{\partial
m^{2}}\bigg|_{m\rightarrow0},\label{eq2}
\end{eqnarray}
It means that the chiral susceptibility measures the response of the chiral condensate to a small perturbation of the current mass of the fermion.

In addition, it is well known that if the dressed fermion propagator has a masslike singularity at complex momenta, instead of a mass singularity on the real timelike axis, it can never be on mass shell and, thus, can never be observed as a free particle~\cite{Phys.Rev.D.22.1452,Phys.Lett.B.217.162,Phys.Rev.C.46.2057,Int.J.Mod.Phys.A.07.5607,Prog.Part.Nucl.Phys.33.477}. In this sense, the absence of a mass singularity
implies directly confinement and, thus, the analytic structure of the dressed fermion propagator might be connected with the confinement. In previous literature, the Euclidean-time Schwinger function is often used to determine whether or not the fermion is confined
\begin{eqnarray}
\Delta(\tau)=\int\textrm{d}^{2}\vec{x}\int\frac{\textrm{d}^{3}p}{(2\pi)^{3}}e^{i(p_{3}\tau+\vec{p}\cdot\vec{x})}\frac{B(p^{2})}
{{A}^{2}(p^{2})p^2+B^{2}(p^{2})},\nonumber\\\label{eq3}
\end{eqnarray}
where the functions ${A}(p^{2})$ and ${B}(p^{2})$ are related to the dressed fermion propagator and denote the fermion wave-function renormalization and the fermion self-energy function, respectively.

Using this Schwinger function, one can show that if there is a stable asymptotic state associated with this propagator, with a mass $\textit{m}$, then
\begin{eqnarray}
\Delta(\tau)\sim e^{-m\tau},\label{eq4}
\end{eqnarray}
for large Euclidean time $\tau$, and so for the logarithmic derivative we get
\begin{eqnarray}
\lim_{\tau\rightarrow\infty}\frac{d\ln\Delta(\tau)}{d\tau}=-m,\label{eq5}
\end{eqnarray}
whereas two complex conjugate masslike singularities, with complex masses $m^{*}$=$a\pm ib$, lead to an oscillating behavior such as
\begin{eqnarray}
\Delta(\tau)\sim e^{-a\tau}\cos(b\tau+\varphi),\label{eq6}
\end{eqnarray}
for large $\tau$.

Because as the number of fermion flavors $N_{f}$ increases, the location of the first oscillations moves to larger values of $\tau$ and it tends to infinity when $N_{f}$ is close to $N_{f}^{c}$, one can identify the reciprocal of the location of the first oscillations as an order parameter for deconfinement phase transition~\cite{Phys.Rev.Lett.77.3724,Few.Body.Syst.46.229}
\begin{eqnarray}
\kappa(N_{f}):=\frac{1}{\tau_{1}(N_{f})},\label{eq7}
\end{eqnarray}
where $\tau_{1}(N_{f})$ signifies the location of the first oscillations of the Schwinger function.

\section{Dyson-Schwinger Equations In The Presence Of Gauge Boson Mass}
The Lagrangian density of QED$_{3}$ with $N_{f}$ flavors of massless fermions in Euclidean space is given by
\begin{eqnarray}
\mathcal{L}=\sum _{i=1}^{N_{f}} \bar{\psi }_i({\not\!\partial} +ie{\not\!A})\psi _i+\frac{1}{4}F_{\mu \nu }^2+\frac{1}{2\xi }(\partial _{\mu }A_{\mu })^2,
\label{eq8}
\end{eqnarray}
where the spinor $\psi_i$ denotes the fermion field with the indices \textit{i}=1,...,$\textit{N}_{f}$ representing different fermion flavors, $A_{\mu }$ signifies the electromagnetic vector field,  $F_{\mu \nu }$ represents the electromagnetic field strength tensor, and $\xi$ is the gauge parameter. With massless fermions, the Lagrangian possesses chiral symmetry and the symmetry group is U(2$N_{f}$). However, when the massless fermion acquires a nonzero mass due to DCSB, the original chiral symmetry will be broken dynamically and the symmetry group reduces to SU($N_{f}$)$\times$SU($N_{f}$)$\times$U(1)$\times$U(1). In (2+1)-dimensional space-time, the lowest rank irreducible representation of the Lorentz group is two-dimensional. In this representation, Dirac fermions are described by two-component spinors and the $\gamma$ matrices may be chosen as the usual Pauli matrices. However, as the three Pauli matrices are a complete set of mutually anticommuting 2$\times$2 matrices, it is impossible to define the other 2$\times$2 matrix that anticommutes with all three $\gamma$ matrices. Consequently, there is nothing to generate a chiral symmetry that would be broken by a mass term $m\bar{\psi}\psi$, whether it be explicit or dynamically generated. Besides, any mass term has the undesirable property that it is odd under parity transformations. Given these, we employ four-component spinors and a four-dimensional representation for the $\gamma$ matrices as in four-dimensional space-time in this paper. A more detail discussion of the reducible and irreducible representations of the Dirac matrices in QED$_{3}$ can be seen in Refs.~\cite{Few.Body.Syst.37.71,J.Phys.A.41.355401}

Based on this Lagrangian density, one can derive the DSEs for the propagators with the help of functional analysis method. The DSE for the dressed fermion propagator is expressed as
\begin{eqnarray}
S^{-1}(p)=S_{0}^{-1}(p)+\Sigma(p),\label{eq9}
\end{eqnarray}
with
\begin{eqnarray}
S_{0}^{-1}(p)=i\gamma\cdot p,\label{eq10}
\end{eqnarray}
and
\begin{eqnarray}
\Sigma(p)=\int\frac{\textrm{d}^{3}k}{(2\pi)^{3}}\gamma_{\mu}S(k)\Gamma_{\nu}(p,k)D_{\mu\nu}(q),\label{eq11}
\end{eqnarray}
where $S^{-1}(p)$ and $S_{0}^{-1}(p)$ are the inverse dressed and free fermion propagators, respectively, $\Sigma(p)$ is the fermion self-energy, $\Gamma_{\nu}(p,k)$ is the dressed fermion-boson vertex, and $D_{\mu\nu}(q)$ is the dressed photon propagator. In addition, based on Lorentz structure analysis, the inverse dressed fermion propagator can be decomposed into
\begin{eqnarray}
S^{-1}(p)=i{\not\!p}A(p^2)+B(p^2),\label{eq12}
\end{eqnarray}
where the functions $A(p^2)$ and $B(p^2)$ are nothing but the functions mentioned in Sec. II. Substituting Eqs. (\ref{eq10}), (\ref{eq11}), and (\ref{eq12}) into Eq. (\ref{eq9}) and taking the traces on both sides of Eq. (\ref{eq9}) after multiplying it with $1$ and $\gamma_{\mu}$, respectively, we arrive at
\begin{eqnarray}
A(p^2)=1-\frac{i}{4p^{2}}\int\frac{\textrm{d}^{3}k}{(2\pi)^{3}} \textrm{Tr}[\gamma\cdot p\gamma_{\mu}S(k)\Gamma_{\nu}(p,k)D_{\mu\nu}(q)],\nonumber\\\label{eq13}
\end{eqnarray}
\begin{eqnarray}
B(p^2)=\frac{1}{4}\int\frac{\textrm{d}^{3}k}{(2\pi)^{3}} \textrm{Tr}[\gamma_{\mu}S(k)\Gamma_{\nu}(p,k)D_{\mu\nu}(q)],\label{eq14}
\end{eqnarray}

For the dressed boson propagator we also have a DSE, namely,
\begin{eqnarray}
D_{\mu\nu}^{-1}(q)=D_{\mu\nu}^{0,-1}(q)+\Pi_{\mu\nu}(q),\label{eq15}
\end{eqnarray}
with
\begin{eqnarray}
D_{\mu\nu}^{0,-1}(q)=q^{2}(\delta_{\mu\nu}-\frac{q_{\mu}q_{\nu}}{q^{2}}),\label{eq16}
\end{eqnarray}
and
\begin{eqnarray}
\Pi_{\mu\nu}(q)= -N_{f}\int\frac{\textrm{d}^{3}k}{(2\pi)^{3}}\textrm{Tr}[\gamma_{\mu}S(k)\Gamma_{\nu}(p,k)S(p)],\label{eq17}
\end{eqnarray}
where $D_{\mu\nu}^{0,-1}(q)$ is the inverse free boson propagator and $\Pi_{\mu\nu}(q)$ is the vacuum polarization tensor. Herein we have chosen the Landau gauge. On the other hand, the vacuum polarization tensor in the presence of gauge boson mass $m_{a}$ has the form
\begin{eqnarray}
\Pi_{\mu\nu}(q)=(q^{2}\Pi(q^{2})+m_{a}^{2})(\delta_{\mu\nu}-\frac{q_{\mu}q_{\nu}}{q^{2}}),\label{eq18}
\end{eqnarray}
where $\Pi(q^{2})$ is the boson vacuum polarization and $m_{a}$ is gauge boson mass which is acquired through the Anderson-Higgs mechanism (a detailed discussion on the Higgs mechanism in QED$_{3}$ can be seen in Refs.~\cite{J.Phys.A.41.255402,Phys.Rev.D.90.073013}). In this paper, we follow
Refs.~\cite{Phys.Rev.B.67.060503,Phys.Rev.D.67.065010,Int.J.Mod.Phys.A.20.2753,Int.J.Mod.Phys.A.24.3969} in adding the gauge boson mass $m_{a}$ by hand and study the effects of it on DCSB and confinement. Substituting Eqs. (\ref{eq16}) and (\ref{eq18}) into Eq. (\ref{eq15}), we obtain the dressed boson propagator
\begin{eqnarray}
D_{\mu\nu}(q)=\frac{\delta_{\mu\nu}-\frac{q_{\mu}q_{\nu}}{q^{2}}}{q^{2}[1+\Pi(q^{2})]+m_{a}^{2}},\label{eq19}
\end{eqnarray}
The vacuum polarization tensor has an ultraviolet divergence, which can be removed by a gauge-invariant regularization scheme. However, this divergence is only present in the longitudinal part, so by contracting $\Pi_{\mu\nu}(q)$ with~\cite{Phys.Rev.D.46.2695}
\begin{eqnarray}
\mathcal{P_{\mu\nu}}=\delta_{\mu\nu}-3\frac{q_{\mu}q_{\nu}}{q^{2}},\label{eq20}
\end{eqnarray}
we can project out the finite vacuum polarization
\begin{eqnarray}
\Pi(q^{2})=\frac{\delta_{\mu\nu}-3\frac{q_{\mu}q_{\nu}}{q^{2}}}{2q^{2}}\Pi_{\mu\nu}(q), \label{eq21}
\end{eqnarray}

It can be found that the coupled DSEs for the fermion and photon propagators form a set of three coupled equations for three scalar functions, and the only unknown function is the dressed vertex function. In principle, we could write down a DSE for the dressed vertex function as well, but this will not lead to a closed set of equations: the DSE for the vertex function involves a four-point function, and so on. The full set of DSEs forms an infinite hierarchy of coupled integral equations for the Green functions. In order to solve the DSE for a particular Green function, we have to truncate or approximate this infinite set of equations. For calculating the propagators, we must find a reasonable approximation for the dressed vertex function. The most simple, and in some sense natural, approximation is to take the leading-order perturbative vertex:
\begin{eqnarray}
\Gamma_{\mu}=\gamma_{\mu},\label{eq22}
\end{eqnarray}
This truncation is usually referred to as rainbow or ladder approximation, since it generates rainbow diagrams in the fermion DSE, and ladder diagrams in the Bethe-Salpeter equation (BSE) for the fermion-antifermion bound state amplitude. Because the rainbow vertex plays the most dominant role in the full vertex in large momentum region and greatly simplifies the coupled integral equations, it is commonly used in studies of the DSEs for fermion and boson propagators. In addition, a range of ansatze for the fermion-boson vertex have been investigated in Ref.~\cite{Phys.Rev.D.70.073007}. There it has been found that the critical number of fermion flavors $N_{f}^{c}$ obtained with the most elaborate construction, obeying the Ward-Takahashi identity, is almost similar to that obtained by the rainbow vertex. Therefore we will study the effects of gauge boson mass $m_{a}$ on DCSB and deconfinement for the rainbow vertex approximation in this work.

Substituting Eqs. (\ref{eq19}) and (\ref{eq22}) into Eqs. (\ref{eq13}) and (\ref{eq14}) and substituting Eqs. (\ref{eq12}), (\ref{eq17}), and (\ref{eq22}) into Eq. (\ref{eq21}), we are then left with
\begin{eqnarray}
A(p^2)&=&1+\frac{2}{p^{2}}\int\frac{\textrm{d}^{3}k}{(2\pi)^{3}}\frac{A(k^2)}{A^{2}(k^2)k^2+B^{2}(k^2)}\nonumber\\
&&\times\frac{1}{q^{2}(1+\Pi(q^{2}))+m_{a}^{2}}\frac{(p\cdot q)(k\cdot q)}{q^{2}},\label{eq23}
\end{eqnarray}
\begin{eqnarray}
B(p^2)&=&2\int\frac{\textrm{d}^{3}k}{(2\pi)^{3}}\frac{B(k^2)}{A^{2}(k^2)k^2+B^{2}(k^2)}\nonumber\\
&&\times\frac{1}{q^{2}(1+\Pi(q^{2}))+m_{a}^{2}},\label{eq24}
\end{eqnarray}
\begin{eqnarray}
\Pi(q^2)&=&\frac{4N_{f}}{q^{2}}\int\frac{\textrm{d}^{3}k}{(2\pi)^{3}}\frac{A(k^2)}{A^{2}(k^2)k^2+B^{2}(k^2)}\nonumber\\
&&\times\frac{A(p^2)}{A^{2}(p^2)p^2+B^{2}(p^2)}\bigg[k^{2}-2(k\cdot q)-\frac{3(k\cdot q)^{2}}{q^{2}}\bigg],\nonumber\\\label{eq25}
\end{eqnarray}
where $q$=$p$-$k$. Here we want to stress that Eq. (\ref{eq24}) has two qualitatively distinct solutions: (a) the Nambu-Goldstone solution, for which $B(p^2)$$\neq$0, describes a phase in which the fermion acquires a nonzero mass and the chiral symmetry is dynamically broken, and (b) the alternative Wigner-Wely solution, for which $B(p^2)$=0, characterises the other phase where the fermion becomes massless and the chiral symmetry is restored.
\section{numerical results}
From the discussions mentioned above, it is obvious that if the momentum dependence of the fermion wave-function renormalization, the fermion self-energy function, and the boson vacuum polarization is obtained, we can further investigate the dynamical fermion mass generation, confinement, and the influences of the gauge boson mass on them. In the following, we will first solve the coupled integral equations numerically for $A(p^2)$, $B(p^2)$, and $\Pi(p^2)$, i.e., Eqs. (\ref{eq23})-(\ref{eq25}).

\begin{figure}[h]
\includegraphics[width=0.45\textwidth]{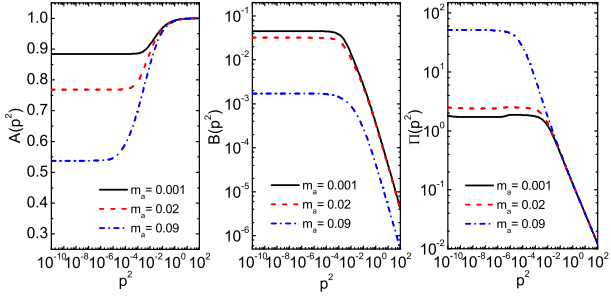}
\caption{Momentum dependence of the functions $A(p^2)$, $B(p^2)$, and $\Pi(p^2)$ for different gauge boson mass $m_{a}$ at $N_{f}$=1.}\label{fig1}
\end{figure}

The three coupled integral equations can be solved numerically by means of iteration method. Starting with a trial function for the fermion self-energy function and the leading-order contribution for the fermion wave-function renormalization, $A(p^2)$=1, we can evaluate the vacuum polarization and solve the coupled equations for $A(p^2)$ and $B(p^2)$. Next, we calculate the vacuum polarization, using these numerical solutions, and iterate this procedure until all three functions converge to a stable solution. The dependence of the functions $A(p^2)$, $B(p^2)$, and $\Pi(p^2)$ on the momentum for several values of the gauge boson mass $m_{a}$ is shown in Fig. \ref{fig1}.

From Fig. \ref{fig1}, we find that the fermion wave-function renormalization is almost constant in the infrared region and approaches one in the ultraviolet region. The infrared constant value of $A(p^2)$ decreases as the gauge boson mass $m_{a}$ increases and the ultraviolet $A(p^2)$ for different $m_{a}$ are all close to one. The fermion self-energy function is nearly constant at small momenta and decreases rapidly to zero at large momenta. With the increasing of the gauge boson mass $m_{a}$, both the infrared and ultraviolet $B(p^2)$ decrease. Also, the boson vacuum polarization is almost constant in the infrared region and decreases rapidly to zero at large momenta. The infrared constant value of $\Pi(p^2)$ increases with the gauge boson mass $m_{a}$ increasing and the ultraviolet $\Pi(p^2)$ for different $m_{a}$ are almost the same.

\begin{figure}[h]
\includegraphics[width=0.45\textwidth]{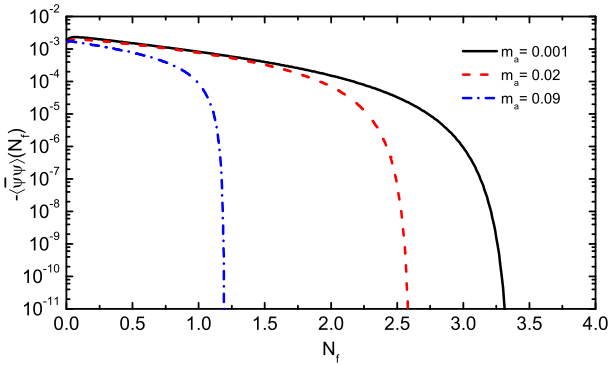}
\caption{The negative of the chiral condensate $-\langle\bar\psi\psi\rangle(N_{f})$ as function of $N_{f}$ for different gauge boson mass $m_{a}$.}\label{fig2}
\end{figure}

For the chiral condensate, substituting Eq. (\ref{eq12}) into Eq. (\ref{eq1}) and then taking the trace, we arrive at
\begin{eqnarray}
\langle\bar\psi\psi\rangle(N_{f})= -4\int\frac{\textrm{d}^3p}{(2\pi)^3}\frac{B(p^2)}{A^{2}(p^2)p^2+B^{2}(p^2)},\label{eq26}
\end{eqnarray}
Because the momentum dependence of the fermion wave-function renormalization and the fermion self-energy function has already been obtained, the chiral condensate can be calculated numerically after substituting the numerical solutions for functions $A(p^2)$ and $B(p^2)$ into Eq. (\ref{eq26}). In Fig. \ref{fig2}, we display the dependence of the chiral condensate as function of the number of fermion flavors $N_{f}$ for several values of the gauge boson mass $m_{a}$.

From Fig. \ref{fig2}, it can be clearly seen that, for a given gauge boson mass $m_{a}$, the negative of the chiral condensate decreases slowly when the number of fermion flavors $N_{f}$ is small and falls rapidly to zero as the $N_{f}$ is close to the critical value. This feature of the chiral condensate shows that the chiral symmetry restoration phase transition in the presence of the gauge boson mass $m_{a}$ is a typical second-order phase transition, which is different from the previous result that the chiral symmetry restoration phase transition without the gauge boson mass $m_{a}$ is a higher-order continuous phase transition~\cite{Phys.Rev.Lett.75.2081}. It is noted that the numerical value of $N_{f}^{c}$ obtained in the present work is much smaller than the previous result obtained by solving the DSE for the fermion self-energy function up to leading-order in 1/$N_{f}$ expansion~\cite{Phys.Rev.B.67.060503}. The comparison of these two results suggests that the fermion wave-function renormalization and the boson vacuum polarization play an important role in the numerical solutions of DSEs and the value of $N_{f}^{c}$. In addition, the critical number of fermion flavors $N_{f}^{c}$ decreases as the gauge boson mass $m_{a}$ increases, which indicates that the gauge boson mass $m_{a}$ weakens the gauge interaction between fermions and, thus, suppresses the occurrence of the DCSB.

For the chiral susceptibility, substituting Eq. (\ref{eq26}) into Eq. (\ref{eq2}) and then performing the derivative, we obtain
\begin{eqnarray}
\chi^{c}(N_{f})&=&4\int\frac{\textrm{d}^3p}{(2\pi)^3}\bigg\{\frac{1}{[A^{2}(p^2)p^2+B^{2}(p^2)]^{2}}\nonumber\\
&&\times\bigg[A^{2}(p^2)B_{m}(p^2)p^2-2A(p^2)B(p^2)A_{m}(p^2)p^2\nonumber\\
&&-B^{2}(p^2)B_{m}(p^2)\bigg]-\frac{1}{p^2}\bigg\},\label{eq27}
\end{eqnarray}
where the functions $A_{m}(p^2)$ and $B_{m}(p^2)$ are just the derivative of the fermion wave-function renormalization and the fermion self-energy function with respect to the current mass of the fermion, respectively, and can be obtained from Eqs. (\ref{eq23})-(\ref{eq25})
\begin{eqnarray}
A_{m}(p^2)&=&-\frac{2}{p^{2}}\int\frac{\textrm{d}^{3}k}{(2\pi)^{3}}\frac{1}{[A^{2}(k^2)k^2+B^{2}(k^2)]^{2}}\nonumber\\
&&\times\frac{1}{[q^{2}(1+\Pi(q^{2}))+m_{a}^{2}]^{2}}\frac{(p\cdot q)(k\cdot q)}{q^{2}}\nonumber\\
&&\times\bigg\{\bigg[\bigg(A^{2}(k^2)k^2-B^{2}(k^2)\bigg)A_{m}(k^2)\nonumber\\
&&+2A(k^2)B(k^2)B_{m}(k^2)\bigg]\bigg[q^{2}\bigg(1+\Pi(q^{2})\bigg)+m_{a}^{2}\bigg]\nonumber\\
&&+A(k^2)\bigg(A^{2}(k^2)k^2+B^{2}(k^2)\bigg)\Pi_{m}(q^{2})q^{2}\bigg\},\label{eq28}
\end{eqnarray}
\begin{eqnarray}
B_{m}(p^2)&=&1+\int\frac{\textrm{d}^{3}k}{(2\pi)^{3}}\frac{1}{[A^{2}(k^2)k^2+B^{2}(k^2)]^{2}}\nonumber\\
&&\times\frac{1}{[q^{2}(1+\Pi(q^{2}))+m_{a}^{2}]^{2}}\nonumber\\
&&\times\bigg\{\bigg[\bigg(A^{2}(k^2)k^2-B^{2}(k^2)\bigg)B_{m}(k^2)\nonumber\\
&&-2A(k^2)B(k^2)A_{m}(k^2)k^2\bigg]\bigg[q^{2}\bigg(1+\Pi(q^{2})\bigg)+m_{a}^{2}\bigg]\nonumber\\
&&-\bigg(A^{2}(k^2)k^2+B^{2}(k^2)\bigg)B(k^2)\Pi_{m}(q^{2})q^{2}\bigg\},\label{eq29}
\end{eqnarray}
\begin{eqnarray}
\Pi_{m}(q^2)&=&\frac{4N_{f}}{q^{2}}\int\frac{\textrm{d}^{3}k}{(2\pi)^{3}}\frac{1}{[A^{2}(k^2)k^2+B^{2}(k^2)]^{2}}\nonumber\\
&&\times\frac{1}{[A^{2}(p^2)p^2+B^{2}(p^2)]^{2}}\bigg[k^{2}-2(k\cdot q)-\frac{3(k\cdot q)^{2}}{q^{2}}\bigg]\nonumber\\
&&\times\bigg\{\bigg(A(p^2)A_{m}(k^2)+A(k^2)A_{m}(p^2)\bigg)\nonumber\\
&&\times\bigg(A^{2}(k^2)k^2+B^{2}(k^2)\bigg)\bigg(A^{2}(p^2)p^2+B^{2}(p^2)\bigg)\nonumber\\
&&-2\bigg[\bigg(A(k^2)A_{m}(k^2)k^{2}+B(k^2)B_{m}(k^2)\bigg)\nonumber\\
&&\times\bigg(A^{2}(p^2)p^2+B^{2}(p^2)\bigg)\nonumber\\
&&+\bigg(A(p^2)A_{m}(p^2)p^{2}+B(p^2)B_{m}(p^2)\bigg)\nonumber\\
&&\times\bigg(A^{2}(k^2)k^2+B^{2}(k^2)\bigg)\bigg]A(p^2)A(k^2)\bigg\},\label{eq30}
\end{eqnarray}

\begin{figure}[h]
\includegraphics[width=0.45\textwidth]{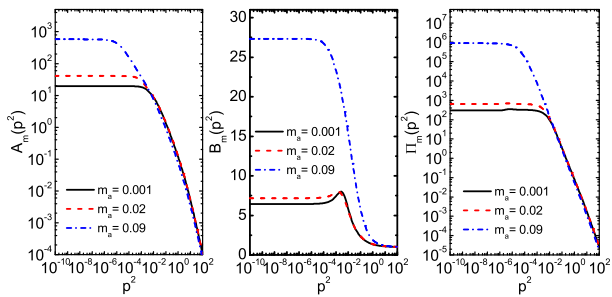}
\caption{Momentum dependence of the functions $A_{m}(p^2)$, $B_{m}(p^2)$, and $\Pi_{m}(p^2)$ for different gauge boson mass $m_{a}$ at $N_{f}$=1.}\label{fig3}
\end{figure}

Since the fermion wave-function renormalization, the fermion self-energy function, and the boson vacuum polarization included on the right-hand side of the above three equations have already been obtained, Eqs. (\ref{eq28})-(\ref{eq30}) are just the coupled integral equations about the functions $A_{m}(p^2)$, $B_{m}(p^2)$, and $\Pi_{m}(q^2)$. We can solve numerically them by employing the iteration method that is used to solve Eqs. (\ref{eq23})-(\ref{eq25}). The dependence of the functions $A_{m}(p^2)$, $B_{m}(p^2)$, and $\Pi_{m}(q^2)$ on the momentum for several values of the gauge boson mass $m_{a}$ is plotted in Fig. \ref{fig3}.

From Fig. \ref{fig3}, we can see that the function $B_{m}(p^2)$ is almost constant in the infrared region and approaches one in the ultraviolet region. The infrared constant value of $B_{m}(p^2)$ increases as the gauge boson mass $m_{a}$ increases and the ultraviolet $B_{m}(p^2)$ for different $m_{a}$ are all close to one. The functions $A_{m}(p^2)$ and $\Pi_{m}(p^2)$ are nearly constant for small momenta and decrease rapidly to zero for large momenta. The infrared constant values of $A_{m}(p^2)$ and $\Pi_{m}(p^2)$ increase with the gauge boson mass $m_{a}$ increasing and the ultraviolet $A_{m}(p^2)$ and $\Pi_{m}(p^2)$ for different $m_{a}$ are all almost the same.

\begin{figure}[h]
\includegraphics[width=0.45\textwidth]{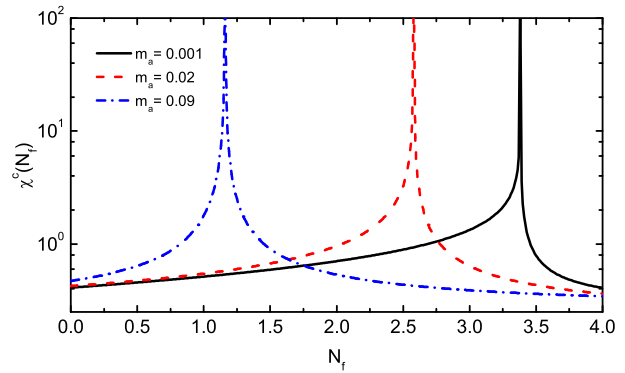}
\caption{The chiral susceptibility $\chi^{c}(N_{f})$ as function of $N_{f}$ for different gauge boson mass $m_{a}$.}\label{fig4}
\end{figure}

As the momentum dependence of the functions $A(p^2)$, $B(p^2)$, $\Pi(p^2)$, and interrelated derivatives has already been obtained, we can substitute the numerical solutions for them into Eq. (\ref{eq27}) and then numerically calculate the chiral susceptibility. In Fig. \ref{fig4}, we depict the dependence of the chiral susceptibility as function of the number of fermion flavors $N_{f}$ for several values of the gauge boson mass $m_{a}$.

From Fig. \ref{fig4}, it is found that, for a given gauge boson mass $m_{a}$, the chiral susceptibility increases slowly when the number of fermion flavors $N_{f}$ is small and exhibits a very narrow, pronounced, and in fact, divergent peak as the number of fermion flavors $N_{f}$ tends to the critical value that is equal to the one obtained by the chiral condensate, which again shows that the chiral symmetry restoration phase transition in the presence of the gauge boson mass $m_{a}$ is a typical second-order phase transition and also indicates that the chiral condensate and the chiral susceptibility are equivalent for characterizing the chiral phase transition. Similarly, the critical number of fermion flavors $N_{f}^{c}$ decreases as the gauge boson mass $m_{a}$ increases, which also reflects that the gauge boson mass $m_{a}$ suppresses the dynamical fermion mass generation.

\begin{figure}[h]
\includegraphics[width=0.45\textwidth]{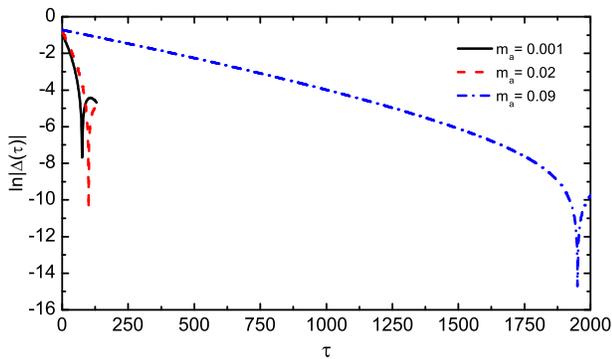}
\caption{The Schwinger function $\ln|\Delta(\tau)|$ as function of $\tau$ for different gauge boson mass $m_{a}$ at $N_{f}$=1.}\label{fig5}
\end{figure}

For the Schwinger function, substituting the numerical solutions for functions $A(p^2)$ and $B(p^2)$ obtained from Eqs. (\ref{eq23})-(\ref{eq25}) into Eq. (\ref{eq3}), we can obtain the behavior of which as function of Euclidean time. The dependence of the Schwinger function on the Euclidean time for several values of the gauge boson mass $m_{a}$ is shown in Fig. \ref{fig5}.

From Fig. \ref{fig5}, we can clearly see that, for a given gauge boson mass $m_{a}$, the Schwinger function exhibits a obvious oscillating behavior as the Euclidean time increases, which indicates that the dressed fermion propagator has a masslike singularity at complex momenta and, thus, the fermion is confined. Here, in order to preserve clarity, we cease plotting each curve at first oscillations in $\ln|\Delta(\tau)|$ but in each case $\ln|\Delta(\tau)|$ exhibits periodic oscillations. With the gauge boson mass $m_{a}$ increasing, the location of the first oscillations moves towards larger values of Euclidean time.

\begin{figure}[h]
\includegraphics[width=0.45\textwidth]{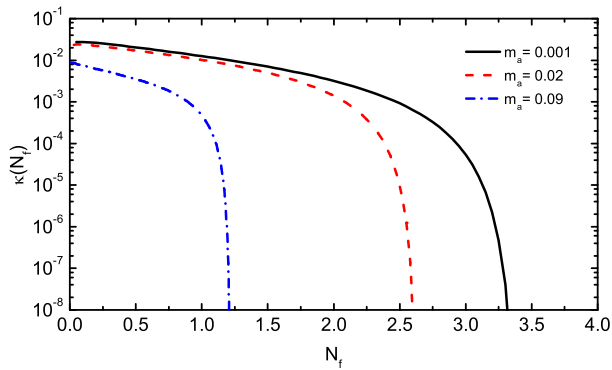}
\caption{The quantity $\kappa(N_{f})$ as function of $N_{f}$ for different gauge boson mass $m_{a}$.}\label{fig6}
\end{figure}

Because the dependence of the Schwinger function as function of the Euclidean time has already been obtained, the quantity $\kappa(N_{f})$ can be obtained via Eq. (\ref{eq7}). In Fig. \ref{fig6}, we display the dependence of the quantity $\kappa(N_{f})$ as function of the number of fermion flavors $N_{f}$ for several values of the gauge boson mass $m_{a}$.

From Fig. \ref{fig6}, it is found that, for a given gauge boson mass $m_{a}$, the function $\kappa(N_{f})$ decreases slowly when the number of fermion flavors $N_{f}$ is small and falls rapidly to zero as the $N_{f}$ approaches the critical value that is equal to the one obtained from the chiral condensate and the chiral susceptibility. The common critical value means that the chiral symmetry restoration and deconfinement phase transitions are simultaneous. The result is different from that reported in the previous literature~\cite{Eur.Phys.J.C.74.3216}. In the case of $N_{f}$$=$1, the authors of Ref.~\cite{Eur.Phys.J.C.74.3216} calculated the Schwinger function with Euclidean time in the region from 0 to 600 for different gauge boson mass $m_{a}$, observed that the logarithm of its absolute value becomes a straight line meaning a stable asymptotic state when the $m_{a}$ exceeds the critical value $m_{a,de}^{c}$$=$0.068 that is smaller than the critical value $m_{a,ch}^{c}$$=$0.1 for the chiral phase transition and, thus, concluded that the occurrence of the deconfinement phase transition is earlier than the chiral phase transition. Actually, from Fig. \ref{fig5}, it can be found that if we calculate the Schwinger function with the Euclidean time in a much larger region, the oscillating behavior of the logarithm of its absolute value remains, even for $m_{a}$$=$0.09. In the present paper, we fellow the Ref.~\cite{Phys.Rev.Lett.77.3724,Few.Body.Syst.46.229} in employing $\kappa(N_{f})$ as the order parameter for deconfinement phase transition and study quantificationally how the quantity $\kappa(N_{f})$ changes with the $N_{f}$ increasing for different $m_{a}$. The basic causal connection for the chiral symmetry restoration and deconfinement phase transitions is a dramatic change in the analytic properties of the propagator which accompanies the disappearance of a nonzero fermion self-energy function. Similarly, the critical number of fermion flavors $N_{f}^{c}$ decreases when the gauge boson mass $m_{a}$ increases, which indicates that the gauge boson mass $m_{a}$ also suppresses the occurrence of the confinement.

\begin{figure}[h]
\includegraphics[width=0.45\textwidth]{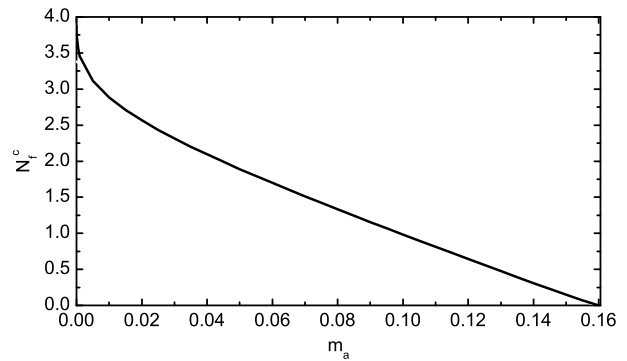}
\caption{The dependence of the critical number of fermion flavors $N_{f}^{c}$ on the gauge boson mass $m_{a}$.}\label{fig7}
\end{figure}

In order to investigate the influences of the gauge boson mass $m_{a}$ on the chiral symmetry restoration and deconfinement phase transitions more complete, we calculate the chiral condensate, the chiral susceptibility, and the function $\kappa(N_{f})$ in a range of gauge boson mass $m_{a}$. The dependence of the critical number of fermion flavors $N_{f}^{c}$ on the gauge boson mass $m_{a}$ is plotted in Fig. \ref{fig7}.

From Fig. \ref{fig7}, we can clearly see that the critical number of fermion flavors $N_{f}^{c}$ decreases gradually as the gauge boson mass $m_{a}$ increases and, thus, there is a boundary that separates the $N_{f}^{c}$-$m_{a}$ plane into two regions. When the number of fermion flavors $N_{f}$ and the gauge boson mass $m_{a}$ are both small, the system is in chiral symmetry breaking/confinement phase, while the system is in chiral symmetry restoration/deconfinement phase as the number of fermion flavors $N_{f}$ and/or the gauge boson mass $m_{a}$ exceed the critical value.

\section{summary and conclusions}
In this paper, based on the experimental observation that there exists a coexisting region between the AF phase and the $\textit{d}$SC phase, we investigate the effects of the gauge boson mass $m_{a}$ on the chiral symmetry restoration and deconfinement phase transitions in QED$_{3}$.

First we solve numerically the coupled integral equations for three scalar functions of the fermion propagator and the boson propagator and discuss the momentum dependence of the three functions and interrelated derivatives with respect to the current mass of the fermion for several values of the gauge boson mass $m_{a}$ at a given number of fermion flavors $N_{f}$.

Then we calculate the chiral condensate, the chiral susceptibility, and the function $\kappa(N_{f})$ within a range of the numbers of fermion flavors $N_{f}$ and the gauge boson mass $m_{a}$. The results show that, for a given $m_{a}$, the chiral condensate and the function $\kappa(N_{f})$ decrease slowly, while the chiral susceptibility increases gradually, when $N_{f}$ is small; As $N_{f}$ approaches the critical value, both the chiral condensate and the function $\kappa(N_{f})$ fall rapidly to zero, which reflects that the chiral symmetry restoration and deconfinement phase transitions are simultaneous and differs from the previous result that the occurrence of the deconfinement phase transition is earlier than the chiral phase transition, whereas the chiral susceptibility exhibits a divergent peak signaling a typically characteristic of second-order phase transition, which is distinct from previous finding that the phase transition without the $m_{a}$ is a higher-order continuous phase transition; In addition, with $m_{a}$ increasing, the $N_{f}^{c}$ decreases, which indicates that the gauge boson mass $m_{a}$ weakens the gauge interaction between fermions and, thus, suppresses the occurrence of the DCSB and the confinement.

Finally, we discuss the relationship between the critical number of fermion flavors $N_{f}^{c}$ and the gauge boson mass $m_{a}$ and find that $N_{f}^{c}$ decreases monotonically with increasing $m_{a}$, which suggests that there exists a boundary separating the $N_{f}^{c}$-$m_{a}$ plane into the chiral symmetry breaking/confinement region for ($N_{f}^{c}$, $m_{a}$) below the boundary and the chiral symmetry restoration/deconfinement region for ($N_{f}^{c}$, $m_{a}$) above the boundary.

\acknowledgments
This work is supported by the National Natural Science Foundation of China (under Grants No. 11275097, No. 11475085, and No. 11535005) and the National Natural Science Foundation of Jiangsu Province of China (under Grant BK20130387).

\bibliographystyle{apsrev4-1}
\bibliography{yin}

\end{document}